# Evaporation kinetics of ferrofluid droplets in magnetic field ambience

**Ankur Chattopadhyay [a], Raghvendra Kumar Dwivedi [a], A R Harikrishnan [b] and Purbarun Dhar [a, c,*]**

[a] Department of Mechanical Engineering, Indian Institute of Technology Ropar, Rupnagar–140001, India

[b] Department of Mechanical Engineering, Birla Institute of Technology & Science Pilani, Rajasthan–333031, India

[c] Department of Mechanical Engineering, Indian Institute of Technology Kharagpur, Kharagpur–721302, India

*<u>Corresponding author</u>:

E–mail: purbarun.iit@gmail.com ; purbarun@mech.iitkgp.ac.in

Phone: +91–1881–24–2119

## Abstract

The present article discusses the physics and mechanics of evaporation of pendent, aqueous ferrofluid droplets and modulation of the same by external magnetic field. We show experimentally and by mathematical analysis that the presence of magnetic field improves the evaporation rates of ferrofluid droplets. First we tackle the question of improved evaporation of the colloidal droplets compared to water, and propose physical mechanisms to explain the same.

Experiments show that the changes in evaporation rates aided by the magnetic field cannot be explained on the basis of changes in surface tension, or based on classical diffusion driven evaporation models. Probing using particle image velocimetry shows that the internal advection kinetics of such droplets plays a direct role towards the augmented evaporation rates by modulating the associated Stefan flow. Infrared thermography reveal changes in the thermal gradients within the droplet and evaluating the dynamic surface tension reveals presence of solutal gradients within the droplet, both brought about by the external field. Based on the premise, a scaling analysis of the internal magneto-thermal and magneto-solutal ferroadvection behavior is presented. The model incorporates the role of the governing Hartmann number, the magneto-thermal Prandtl number and the magneto-solutal Schmidt number. The analysis and stability maps reveal that the magneto-solutal ferroadvection is the more dominant mechanism, and the model is able to predict the internal advection velocities with accuracy. Further, another scaling model to predict the modified Stefan flow is proposed, and is found to accurately predict the improved evaporation rates.

**Keywords:** Evaporation; pendant droplet; ferrofluid; ferro-hydrodynamics; Marangoni advection, particle image velocimetry; magnetic field

## 1. Introduction

Evaporative hydrodynamics, thermal and species transport in droplets involves rich and intriguing physics, and is of importance to a wide bevy of applications. Examples of such applications are in the automobile industry (fuel injection and spray) and propulsion systems [1-3], in combustors [4], two-phase systems [5, 6], in HVAC components, etc. Evaporation kinetics of droplets is also of importance to pesticide fumigation [7], ink-jet printing, and manufacturing technologies and microscale lithography [8], etc. Hence understanding of the hydrodynamics heat, and mass transport associated with evaporation kinetics of droplets holds importance for design and development of such systems.

Typically, droplets can be categorized into two types: pendant and sessile. For understanding evaporation kinetics, the pendant droplet approach has been adapted since such drops resemble free-standing droplets very closely, and unlike sessile droplets, their evaporation behavior is independent of surface features [12-14]. Godsave [15] initially investigated the evaporation behavior of near-spherical droplets and reported that molecular diffusion is the major driving parameter behind the evaporation kinetics. The $D^2$ law was introduced as a mathematical representation to describe the temporal variation of the diameter of the evaporating droplet. This observation was later confirmed by Kuz [16] and has since been a robust yet simple mathematical description for pendant droplet evaporation kinetics.

Multicomponent systems have also been studied to understand the transport processes during evaporation or vaporization [14, 17, 18]. Among such multicomponent systems, colloidal complex fluids have received attention, since their thermophysical transport properties can lead to manipulation and altering of evaporation behavior of the fluid [9-11]. In a colloidal complex fluid, the presence of the dispersed phase can alter the interfacial properties [19, 20], which can affect the evaporation dynamics of the droplets. The evaporation of such multicomponent droplets presents complicated physics as the species transport is function of the diffusivities, or interactions, or concentrations of constituent species. Gerken et al. [21] reported that enhancing the concentration of dispersed particles leads to reduced evaporation rates of the colloids. It was argued that the agglomeration of the nanoparticles at the liquid-vapor interface of the droplet reduces the liquid fraction available for evaporation at the interface, leading to diminished evaporation rate. Chen et al. reported that the addition of particle changes the latent heat of vaporization of the colloidal system and changes the evaporation rate [22]. Harikrishnan et al. investigated the evaporation kinetics of complex fluids containing surfactants, nanoparticles, etc. [23]. The interplay of internal advection, Marangoni effect and Brownian transport of particles were shown to be responsible for the enhancement of evaporation rates.

The transport phenomena in such complex fluid droplets can be tuned and modified as needed by the use of external stimulus, such as electromagnetic, thermal, optical or acoustic fields. Investigations by Fattah et al. have shown that complex microstructures in an elastomer matrix can be printed by controlling the motion of ferrofluid droplet [24]. Besides, a few studies have also examined the possibilities of remote actuation of ferrofluid droplets by a combination of uniform and non-uniform magnetic field strength [25, 26]. It has been shown that electromagnetic fields can induce a surface energy gradient, which can modify the interfacial tension gradient on the liquid, thus enabling actuation of a droplet along surface [27-29]. Havard et al. reported the breakup of a ferrofluid pendant droplet under the influence of a horizontal magnetic field and discussed the evolution of droplet shapes due to the field effects [30]. Droplet evaporation of paramagnetic solutions under the influence of magnetic field shows that magneto-solutal advection within the droplet leads to augmented evaporation [31].

The present article focuses on the physics of evaporation of ferrofluids in the presence of magnetic field. Pendant droplets have been considered to eliminate the role played by the wetting interactions of the droplet with the substrate and focus on the role of the magnetic field. Experiments are performed to understand the roles played by the magnetic properties of the ferrofluid, concentration of the magnetic nanoparticles, and the magnetic field intensities in altering the evaporation kinetics. The role of the surface tension and diffusion driven evaporation behavior under field effect have been explored. Particle image velocimetry (PIV) has been done to qualitatively and quantitatively understand the nature of internal hydrodynamics within such droplets and their role in modulating the evaporation process. The internal hydrodynamics is proposed to be borne of the ferro-advection in the ferrofluid due to the magnetic field environment. An analytical scaling analysis has been proposed to determine the dominant governing parameter responsible for such changed evaporation behavior. A scaling model to predict the modified evaporation rates has also been proposed. The findings may have important implications in microhydrodynamics systems with magnetic components and actuation.

## 2. Materials and methodologies

Three different magnetic nanoparticles (MNPs) - $Fe_2O_3$, $Fe_3O_4$, and NiO (the first two procured from Alfa Aesar, India and the last from Nanoshel, USA) have been used to synthesize the ferrofluids. Figure SF1 (refer supplementary information) illustrates the High-Resolution Scanning Electron Microscopy (HRSEM) images of different MNPs and discusses their morphologies. Fig. SF2 (supplementary information) illustrates the magnetization curves (M-H) of the MNPs at 300 K and discusses the magnetic properties of the particles. Anhydrous MNPs were dispersed in deionized water and were mechanically stirred for 1 hour. Next, the ferrofluids were ultra-sonicated for 2–3 h (Oscar Ultrasonics, India) to form homogeneous complex fluids.

No macroscale agglomeration or sedimentation was noticed in the ferrofluids for 1-2 days, which is significantly large compared to the experimental time span. The ferrofluids were also visibly stable when subjected to uniform magnetic fields up to ~ 0.24 T during the experiments.

Figure 1 illustrates the experimental setup used in the present study. The pendant droplets were generated using a precision syringe pump (New Era Pump Systems Inc., USA) connected to a flexible drip tube, with a 27 gauge needle at the other end. The pendant droplet (diameter ~2.8 ± 0.2 mm) is suspended from the tip of the needle. The droplet was positioned at the center of the poles of an electromagnet (Holmarc Optomechantronics, India) which was controlled by a digitized current source (Polytronic Corporation, India). A precision gaussmeter was used to initially calibrate the magnetic field strength at the center of the poles with respect to the applied current. Three field strengths, viz. 0.08, 0.16 and 0.24T have been studied in the present case. At fields of 0.3 T or above, the pole shoes exhibit mild heating, which will modify the evaporation rates, and hence the experiments were limited to 0.24 T.

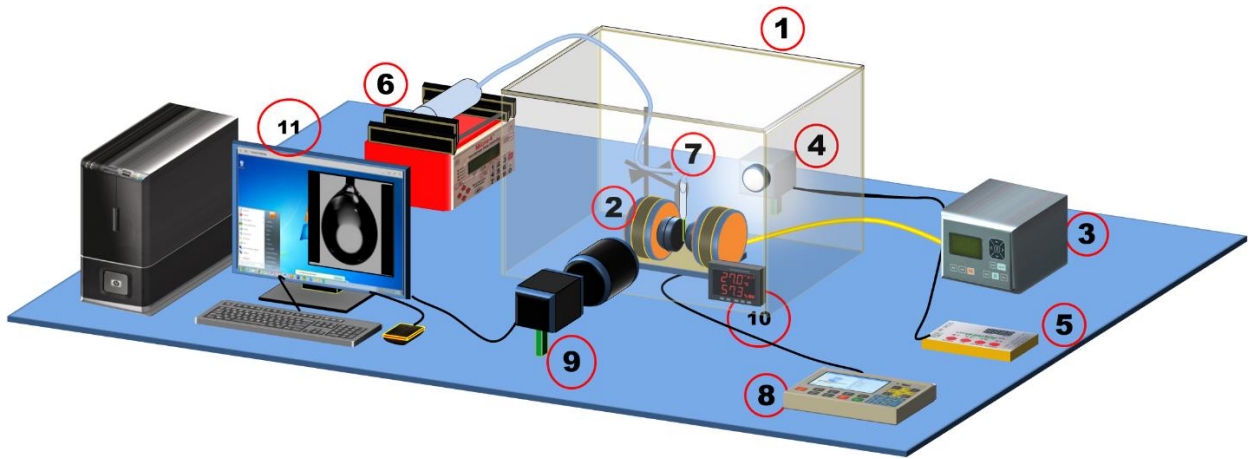

**Figure 1:** Schematic of the experimental setup (1) acrylic chamber, (2) electromagnet with gaussmeter, (3) power supply to electromagnet, (4) LED array for backlight illumination, (5) LED array controller, (6) micro-liter syringe pump, (7) flexible tubing and needle system to generate pendant droplet, (8) gaussmeter control unit, (9) CCD camera with long distance microscope lens, (10) digital thermometer and hygrometer, (11) computer system for camera control and data acquisition.

The droplet and electromagnet assembly was placed within an acrylic enclosure to prevent ambient disturbances. The temperature and humidity in the chamber (measured ~ 10 mm away from the droplet with a probe) were noted as 24 ± 2 $^{o}$C and 53 ± 3 % during all the experiments. A charge coupled device (CCD) camera (Holmarc Opto-Mechatronics, India) with a long-distance microscope lens was used to image the evaporation process. A diffused (to suppress direct radiant heating of the droplet) LED array was used as the backlight. The images were analyzed in the open-source processor ImageJ. Infrared imaging of the evaporating droplets has been done to assess the thermal state of the droplet under magnetic field. The thermal camera (FLIR T650sc) has 640 × 480 pixel array with a thermal sensitivity of ± 0.02 °C at 30°C, and a spectral range of 7.5 - 14 μm.

Flow visualization and quantification within the droplet was done using particle image velocimetry (PIV). Neutrally buoyant fluorescent seed particles (polystyrene, diameter ~10 μm, Cospheric LLC, USA) were dispersed in the dilute $Fe_2O_3$ ferrofluids (for visualization with minimal noise). A continuous wave laser (532 nm wavelength, maximum power 10 mW, Roithner GmbH, Germany) has been used as the illumination source. A cylindrical lens was used to generate a light sheet (~ 0.5-0.75 thick), which was focused at the vertical mid-plane of the droplet. The PIV was done during the first 5 minutes of the evaporation process to eliminate artefacts due to change in ferrofluid concentration. The PIV was done at 20 fps, with a camera resolution of 2048 x 2048 pixels, for 60 seconds, and with interrogation resolution of ~ 200 pixels/mm. The images were post-processed using the open-source PIV-lab [32, 33]. Standard noise suppression algorithms were used to obtain better SNR. Standard intensity capping algorithms were used to improve the intensity of the seeding pixels. A 4 pass cross-correlation scheme was employed, with consecutive interrogation windows of 64, 32, 16 and 8 pixels. Correlation coefficients of ~ 0.75 were ensured during the post-processing.

## 3. Results and discussion

### 3.A. Evaporation kinetics of ferrofluids under magnetic field constraint

Figures 2(a) and 2 (c) illustrate the variation of the non-dimensional square of the droplet diameter ($\psi = \frac{D^2}{D_0^2}$) for 0.1 wt. % $Fe_2O_3$ and 0.01 wt. % $Fe_3O_4$ ferrofluids with the time scale ($\tau = \frac{t}{D_0^2}$). $D_0$ represents the initial diameter of the droplet and $D$ is the instantaneous droplet diameter. Since only the water evaporates from the droplet, the $D^2$ law (eqn. 1) (Abramzon and Sirignano [34]) holds true.

$$\psi = 1 - k\tau \quad (1)$$

In eqn. 1, $k$ is the evaporation rate constant. It can be observed from fig. 2 that the ferrofluids evaporate faster than the water droplet. With the application of the magnetic field, the increment in the evaporation rate further enhances, as a direct function of the field strength.

The variations of the evaporation rate constant ($k$) due to magnetic field stimulus for $Fe_2O_3$ and $Fe_3O_4$ ferrofluids are illustrated in figs. 2 (b) and 2 (d). The evaporation kinetics of NiO ferrofluids has been illustrated in fig. 3 (a). Comparison of the augmented evaporation rates of different ferrofluids (all 0.05 wt. %) due magnetic field has been shown in fig. 3 (b). Upon drawing parallels between fig. 3 (b) and fig. SF2 (supporting information), it may be inferred that improvement of evaporation rate under magnetic field effect is directly proportional to the magnetization of the constituent particle phase. This brings to the forefront the possible role of Lorentz forces on the ferrofluid on the modulation of the evaporation behavior. With an understanding of the nature in which ferrofluids may evaporate faster compared to the base fluid, the next component shall be to discuss the mechanism of further improved evaporation due to external magnetic fields.

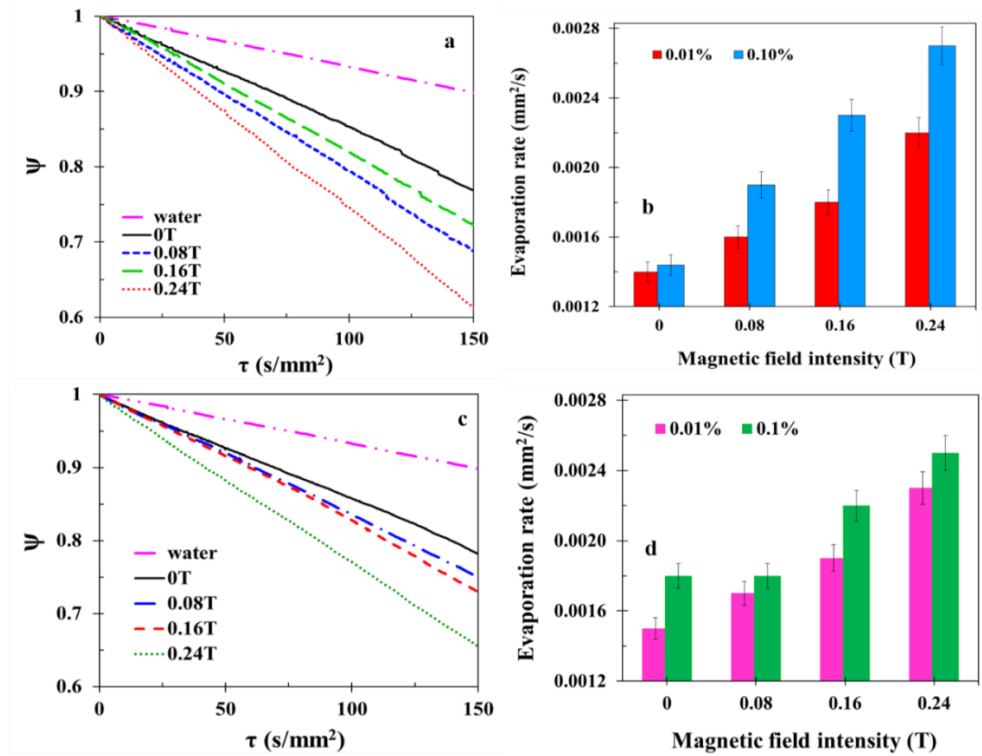

**Figure 2:** (a) Evaporation characteristics of $Fe_2O_3$ ferrofluids (0.1 wt. %), (b) comparison of evaporation rates of different concentrations (0.01 and 0.1 wt. %) of $Fe_2O_3$ ferrofluids at different magnetic field intensity, (c) same as (a) but for $Fe_3O_4$ ferrofluids (0.01 wt. %), (d) same as (b) but for $Fe_3O_4$ ferrofluids

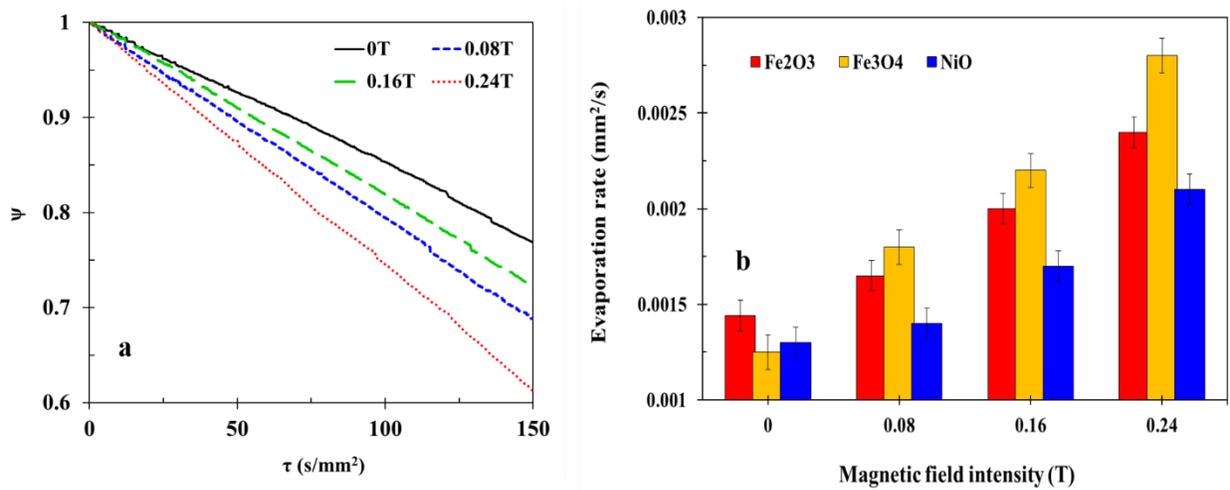

**Figure 3:** (a) Evaporation characteristics of NiO ferrofluids (0.05 wt. %) (b) comparison of evaporation rates of different ferrofluids (0.05 wt. %) as function of magnetic field intensity

**3.B. Role of surface tension and diffusion driven evaporation**

Fig. 4 (a) illustrates the surface tensions (STs) of the ferrofluids at different magnetic fields. The STs have been determined using the pendant drop shape analysis method. The STs improve by minor amounts with the addition of particles, and it is caused by the interfacial adsorption–desorption characteristics of the nanoparticles at the bulk and interface of the droplet [11, 19]. Increasing the particle concentration leads to further minor increase in the STs (fig. SF3, refer supplementary). Minor increase in STs due to nanoparticles signifies reduced propensity of evaporation as a larger surface energy barrier is to be overcome by the evaporating molecules. This is in contrast to the observations of enhanced evaporation. Additionally, the external magnetic field is noted to have no significant effect on the STs. Consequently, the evaporation kinetics of the ferrofluids cannot be explained by appealing to the changes in STs.

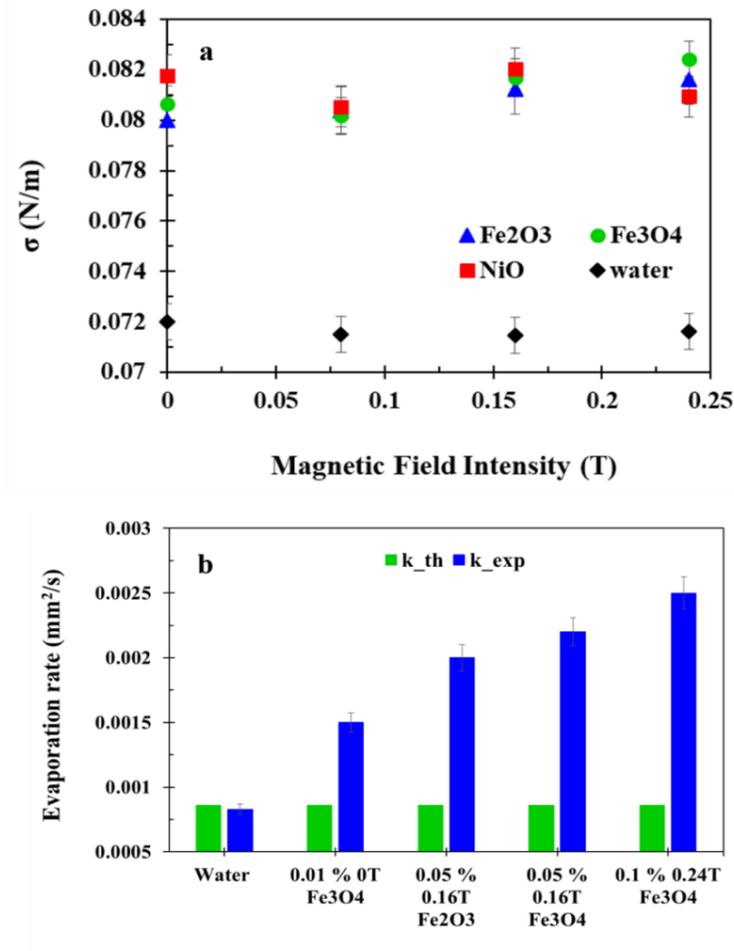

**Figure 4:** (a) Surface tension of water and 0.05 wt. % ferrofluids under magnetic field (b) comparison between experimental (k_exp) and theoretically predicted (k_th) evaporation rates.

The contribution of diffusion driven evaporation behavior towards the enhanced evaporation is investigated. The theoretical diffusion-driven evaporation rate proposed by Abramzon and Sirignano [34] (refer supplementary for model details) has been compared with the experimental values in fig. 4 (b). The theory considers the diffusion of the water vapor molecules from the diffusion layer surrounding the droplet to the far field by virtue of the vapor concentration gradient. Thereby, the model agrees well with the observations for the simple water case. However, the model fails to predict the evaporation rates for the ferrofluids, both in the absence and presence of field. Since the model is deduced only for the vapor side conditions, the predictions show that there are no changes apparent on the vapor side, and hence the internal behavior of the droplets requires to be probed.

## 3.C. Internal hydrodynamics of the droplets

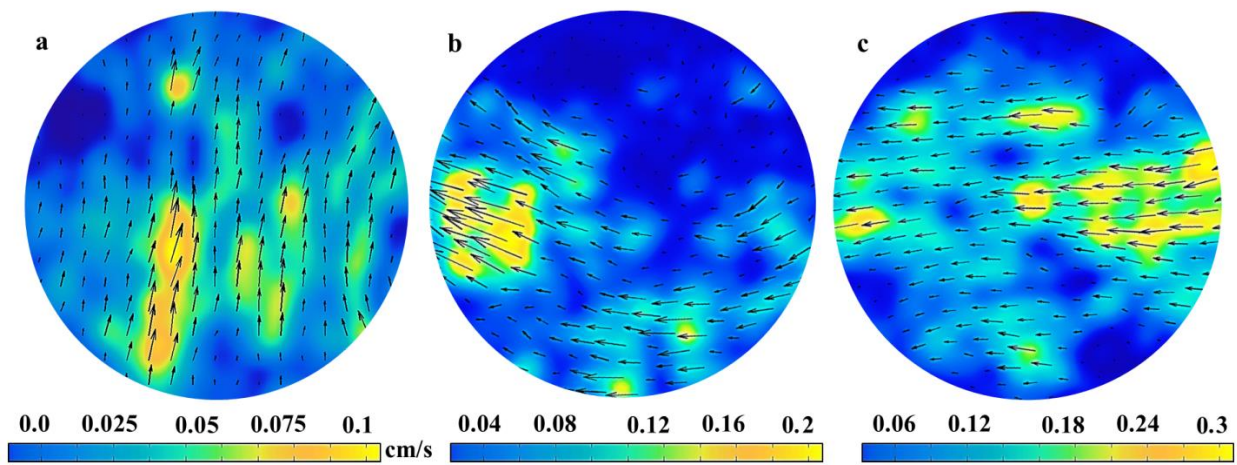

**Figure 5.** Velocity contours and vector field within 0.01 wt. % $Fe_2O_3$ ferrofluid droplets for different field constraints, (a) 0T (b) 0.08T and (c) 0.16T. Only the spherical bulb region of the pendant has been analyzed and the neck region is not shown. All velocities are in cm/s.

The internal hydrodynamics of the ferrofluid droplets during evaporation has been quantified (using PIV) to understand its role, if any, on the improved evaporation. Only $Fe_2O_3$ ferrofluids (0.01 wt. %) were tested as other ferrofluids posed the problem of opacity. The PIV studies were carried out during the initial 5 minutes of evaporation to ensure that change in particle concentration does not affect the velocimetry. The velocity contours and vector fields obtained from post-processing are illustrated in fig. 5. The water droplets (not illustrated) do not exhibit any internal advection except for minor and occasional drift motion, which is in agreement with literature [14, 23].

In the case of the ferrofluid droplets, (zero field), weak, albeit well-defined advection pattern is noticeable (fig. 5 (a)). In the presence of the magnetic field, the average velocity of advection is noted to enhance and the internal flow is well-defined and consistent with time. In the absence of field the flow is directed from the bulb of the pendent towards the neck (fig. 5 (a)). In the field environment (in fig. 5, the field is directed horizontally across the droplet), the direction of advection changes and flow occurs as shown in figs. 5 (b) and (c). This change in direction of advection of a conducting paramagnetic fluid is consistent with the description of Fleming's right hand rule [31, 35]. Hence it is seen that the evaporation of the ferrofluid droplet in field environment is conjugated to increase in internal advection. Similar to the discussion on the improved evaporation of colloids, the circulation promotes higher shear at the liquid-vapor interface shrouding the droplet. The shear entrains ambient air within the vapor diffusion layer, constantly replenishing it, and leading to higher evaporation rate [23, 31]. However, the genesis of this improved advection required further analysis and probing.

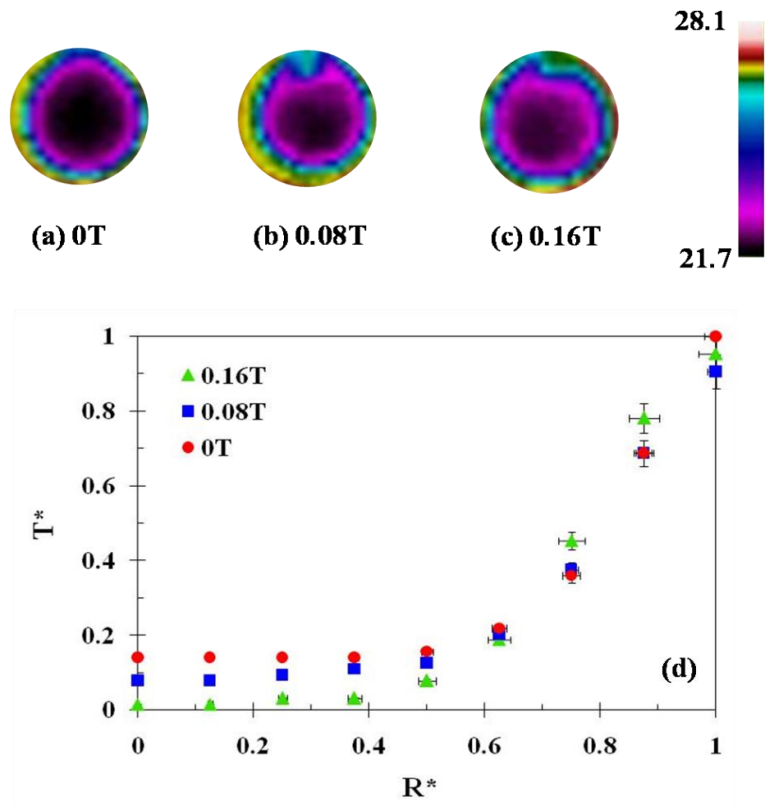

**Figure 6:** Infrared images of the evaporating ferrofluid droplets (0.1 wt. % $Fe_2O_3$) at, (a) 0T, (b) 0.08T, (c) 0.16T, (d) radial distribution of non-dimensional temperature in the droplet. Here

$R^* = \dfrac{r}{R}$, r=radial location, R=radius of droplet, and $\left(T^* = \dfrac{T_r - T_{r=0}}{T_{r=R} - T_{r=0}}\right)$. The temperature scale for (a), (b) and (c) are in °C.

Fig. 6 illustrates the infrared thermography images of 0.1 wt. % $Fe_2O_3$ ferrofluid droplets under 0T, 0.08T, and 0.16T fields. A darker central zone appears in the zero field case. With increasing field magnitude, the relative darkness reduces, thereby indicating a higher value of temperature around the core of the droplet. Under the magnetic influence, the central portions of the drop acquire higher temperatures than in absence of field. Thereby the magnetic environment increases the radial thermal gradients within the droplet. The radial variations of the non-dimensional temperatures within the droplet have been analyzed from the infrared images and are shown in fig. 6 (d). It is noted that while the thermal profiles towards the droplet surface are similar in nature, there is considerable difference within the inner parts of the droplet. This signifies that the field modulated evaporation of the ferrofluid droplet also triggers a thermal gradient change within the droplet. If a droplet evaporates faster, the non-dimensional thermal profile within the droplet must remain same as a slow evaporating droplet, under similarly steady conditions. The change in the thermal distribution illustrates that in addition to augmented evaporation kinetics, the ferrofluid droplets within magnetic field environment also exhibit differences in the thermal advection patterns within the droplet. Hence, analysis of the magneto-thermal advection within the droplet is essential to probe deeper.

### 3.D. Role of magneto-thermal ferro-advection

The internal hydrodynamics in the droplet may be generated by two possible factors in the present case, viz. thermal and solutal gradients within the droplet and along the droplet interface. The shape of the pendant drop is known to induce non-uniform evaporation rates along the droplet surface [23, 35-37], thereby leading to thermal gradients and thereby thermal Marangoni advection. Additionally, the evaporation induced cooling of the droplet leads to thermal gradients across the droplet, leading to internal thermal advection. The thermal advection and its behavior in magnetic field could be a possible responsible mechanism behind the internal ferrohydrodynamics.

A mathematical scaling analysis (refer supplementary for detailed derivation) is developed to understand the role of the thermal advection, in the presence and absence of magnetic field. The analysis has been proposed based on similar propositions in literature [14, 23]. The energy balance expression of an evaporating ferrofluid drop in a magnetic field environment is as

$$\dot{m}h_{fg} = \frac{k_{th}A\Delta T_m}{R} + \rho C_p A u_{c,m} \Delta T_m + \rho C_p A u_f \Delta T_m \tag{2}$$

where $\dot{m}$, $h_{fg}$, $\rho$, $C_p$, $k_{th}$, and $R$ denote the mass loss due to evaporation, enthalpy of vaporization, density, specific heat and thermal conductivity and instantaneous radius of the ferrofluid droplet, respectively. $A$, $\Delta T_m$, $u_{c,m}$, and $u_f$ represent the effective surface area of evaporation, the temperature difference between the droplet's center and surface, internal circulation velocity due to pure thermal advection and the magnetic field induced circulation velocity, respectively. The left hand side of eqn. 2 represents the energy corresponding to evaporative mass flux, which is balanced by the sum of the heat diffusion across the droplet, the thermal advection contribution, and component generated due to the magneto-thermal advection.

The internal velocity of thermal advection at zero field is expressed as

$$u_{c,m} = \frac{\sigma_T \Delta T_m}{\mu} \tag{3}$$

where, $\sigma_T$ is the rate of change of ST due to temperature and $\mu$ is the viscosity of the ferrofluid. The magneto-thermal advection velocity is expressed in terms of the governing Hartmann number $\left(Ha = \frac{BMR\rho}{u_{c,m}\mu}\right)$ as [31]

$$u_f \sim Ha\sqrt{\frac{\sigma_T \Delta T_m}{\rho R}} \tag{4}$$

Substituting the $u_{c,m}$ and $u_f$ in eqn. 2, the expression becomes

$$\rho \dot{R} R h_{fg} = k_{th}\Delta T_m + \rho C_p R \Delta T_m \left(\frac{\sigma_T \Delta T_m}{\mu} + Ha\sqrt{\frac{\sigma_T \Delta T}{\rho R}}\right) \tag{5}$$

Introducing the thermal Marangoni number $\left(Ma_T = \frac{R}{\alpha}\sqrt{\frac{\dot{R} h_{fg}\sigma_T}{\mu C_p}}\right)$ in eqn. 6 and algebraic manipulation yields [31]

$$\rho \dot{R} R h_{fg} = k_{th}\Delta T_m \left(1 + Ma_T + Ha\sqrt{Ma_T \Pr}\right) \tag{6}$$

where $Pr$ is the Prandtl number. For stable internal circulation $Ma_T \sim 80$, and thus $1+ Ma_T \sim Ma_T$. Therefore, the eqn. 6 is modified as

$$\frac{\rho \dot{R} R h_{fg}}{k_{th}\Delta T_m} = Ma_T + \sqrt{HaMa_T}\sqrt{Ha\Pr} = Ma_T + \sqrt{Ma_{T,m}\Pr_m} \tag{7}$$

where, $Ma_{T,m}$ and $Pr_m$ represent the magneto-thermal Marangoni number and the magnetic Prandtl number respectively.

Fig. 7 illustrates the percentage contributions of each component (eqn. 2) for a 0.1 wt. % $Fe_2O_3$ ferrofluid droplet. The thermal diffusion has negligible contribution, while the dominant mode of thermal transport is due to thermal advection. This is in agreement with colloidal evaporation kinetics [23]. With the introduction of the magnetic field, the magneto-thermal advection component appears. As the field strength increases, the magneto-thermal component increases in magnitude at the expense of the thermal-advection component. This signifies that a significant portion of the energy transport mode (~ 30 % at highest field strength) is governed by the magneto-thermal advection within the ferrofluid droplet. Thereby magneto-thermal advection could be a possible genesis for the observed improvement in internal advection strength.

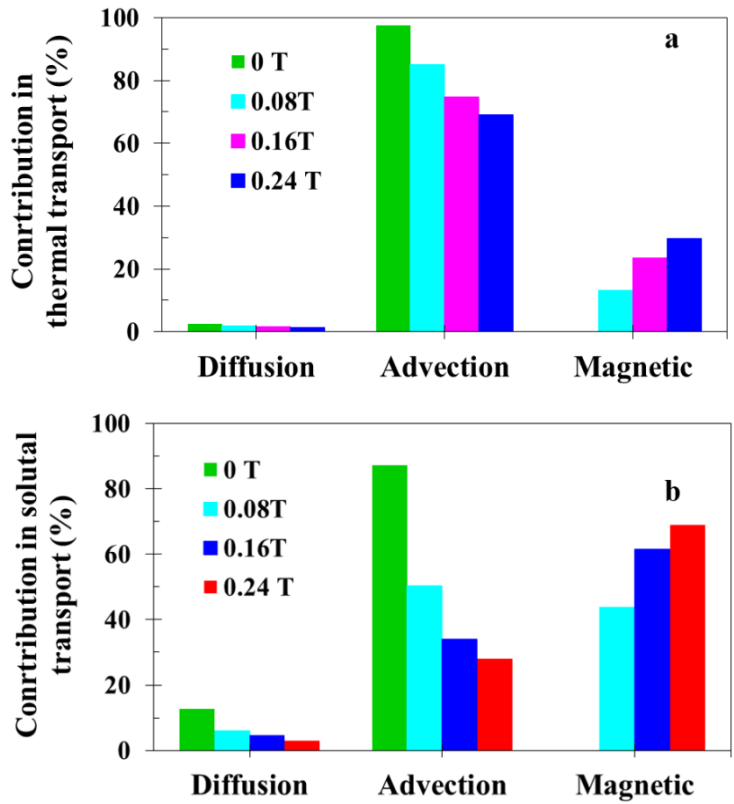

**Figure 7:** (a) Contribution of different mechanisms in the magneto-thermal advection model (0.1 wt. % $Fe_2O_3$ ferrofluid), (b) contribution of different mechanisms in the magneto-solutal advection model (0.1 wt. % $Fe_2O_3$ ferrofluid).

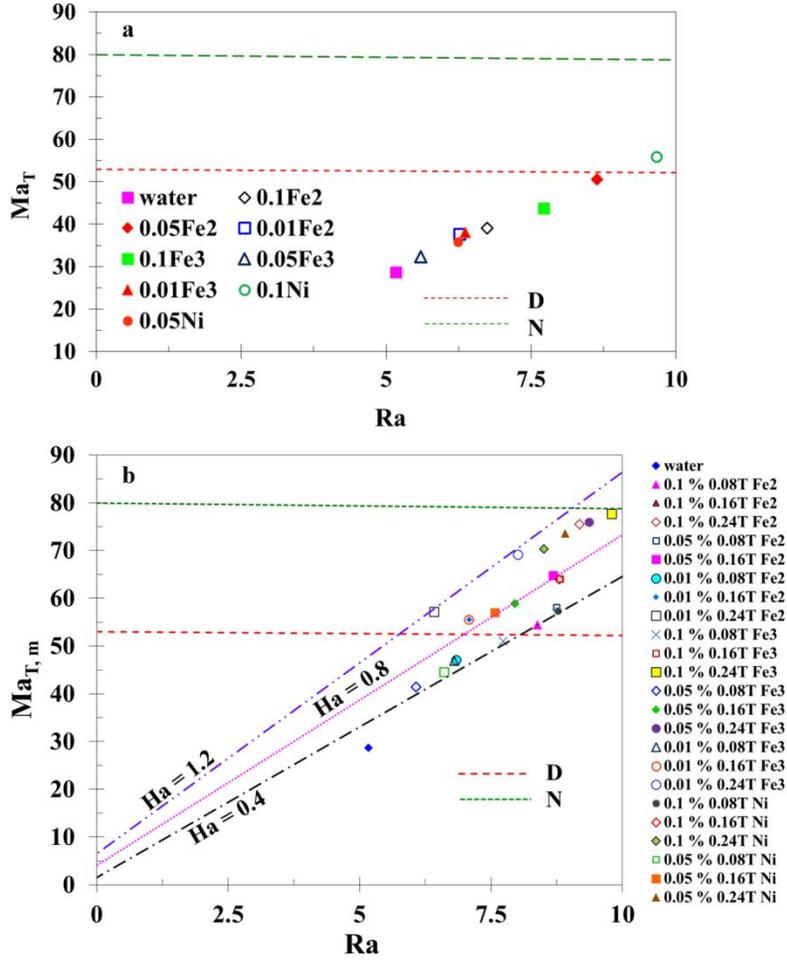

**Figure 8:** (a) Phase plot of the Rayleigh and thermal Marangoni numbers for different ferrofluid droplets evaporating in the absence of magnetic field, (b) comparison of magneto-thermal Marangoni and Rayleigh number for different ferrofluid droplets evaporating in presence magnetic field. 'N' and 'D' lines represent the stability criteria proposed by Nield [38] and Davis [39], respectively.

Thermal advection can be generated by the thermal Marangoni effect, or the buoyance driven Rayleigh effect within the droplet. The circulation velocity only due to Rayleigh advection within the droplet is expressed as [14]

$$u_{c,r} = \frac{\rho g \beta \Delta T_r R^2}{\mu} \tag{8}$$

The Rayleigh number and the temperature difference driving the buoyant advection are expressed as

$$Ra = \frac{R^2}{\alpha}\sqrt{\frac{\rho g \beta \dot{R} h_{fg}}{\mu C_p}} \qquad (9)$$

$$\Delta T_r = \sqrt{\frac{\mu \dot{R} h_{fg}}{\rho g \beta R^2 C_p}} \qquad (10)$$

In eqns. 8-10, $g$, $\beta$, and $\Delta T_r$ represent the acceleration due to gravity, coefficient of thermal expansion of the fluid, and the temperature difference within the droplet driving the Rayleigh advection [23].

Since both $Ma_T$ and $Ra$ based advection is possible within the droplet, the dominant mode (if any) needs to be determined. According to the stability criterion proposed in literature [38], for a Marangoni-Rayleigh advection system,

$$\frac{Ra}{Ra_c} + \frac{Ma}{Ma_c} = 1 \qquad (11)$$

where, the critical Rayleigh and Marangoni numbers (for the onset of stable advection) are indicated by subscript 'c'. Fig. 8 (a) illustrates a phase plot between the $Ma_T$ and the $Ra$ for the ferrofluids at zero field condition. The $Ra_c$ is ~ 1708, as per Chandrasekhar's seminal stability theory. The stability condition for the $Ma_C$ (Davis [39]) based on the energy theory approach shows the sufficient condition for stability on internal advection. The criteria (Nield [38]) however, derived from the linear stability theory, provides the paradigm of unconditional stability of internal thermal advection. As $Ma_c$ is ~ 55 and ~ 81 as per the Davis and Nield criteria, respectively. Hence the probability of Rayleigh advection within the droplets being the dominant reason for circulation is minimal, and the Marangoni advection is the more potent mechanism (as the critical Marangoni value is much less compared to the critical value for Rayleigh advection). It is observed from fig. 8 (a) that the presence of the particles in the ferrofluids leads to improved $Ma_T$, however, the points lie below the regions of intermittent stable circulation. This is in agreement with observations as the ferrofluid droplets show minor advection behavior of intermittent nature.

Fig. 8 (b) illustrates a similar phase plot, but for ferrofluids droplets under the influence of magnetic field. Consequently, the effective magneto-thermal Marangoni number ($Ma_{T,m}$) has been used. To understand the role of the ferro-advection behavior, iso-Ha lines [40] have been incorporated in the phase plot. A large number of points now lie within the regime of partially stable circulation, between the Davis and Nield criteria. This signifies that the droplet exhibits partially stable internal advection behavior, which is in agreement with the ferro-advection behavior observed in the presence of magnetic field. The iso-Ha lines show that with increasing $Ha$, the points shift towards the stable advection regime. This signifies that the strength of thermal-ferro-advection enhances with the magnetic field parameters. However, it cannot be inferred at this moment if the magneto-thermal behavior is the main governing mechanism behind the ferro-advection.

## 3.E. Role of magneto-solutal ferro-advection within the droplet

During the evaporation, only the water vaporizes and with time, the particle concentration of the ferrofluid enhances. As the presence of nanoparticles improves the surface tension of the fluid, it signifies that the particles preferentially desorb away from the interface compared to the bulk [19]. Thereby a ferrofluid droplet exhibits an innate particle concentration gradient between the bulk and the interface, which drives the solutal advection. Further, the non-uniform evaporation from the prolate shape leads to concentration gradient at the interface itself, leading to solutal Marangoni advection. The nanoparticle concentration in the bulk is deduced from the instantaneous volume of the droplet (the product of the volume of droplet and bulk concentrationwill remain constant, CV=constant). The instantaneous ST of the ferrofluid is obtained by fitting the droplet shape to the Young-Laplace equation. The ST tension of the droplets as function of concentration and magnetic field are known from carefully designed experiments. The transient ST data is mathematically correlated to this dataset, and upon elimination of the ST component, the transient interfacial concentration of the nanoparticles is obtained [31]. Figure 9(a) illustrates the transient bulk and the interfacial concentrations for 0.1 wt. % $Fe_2O_3$ ferrofluid droplet for different magnetic field strengths.

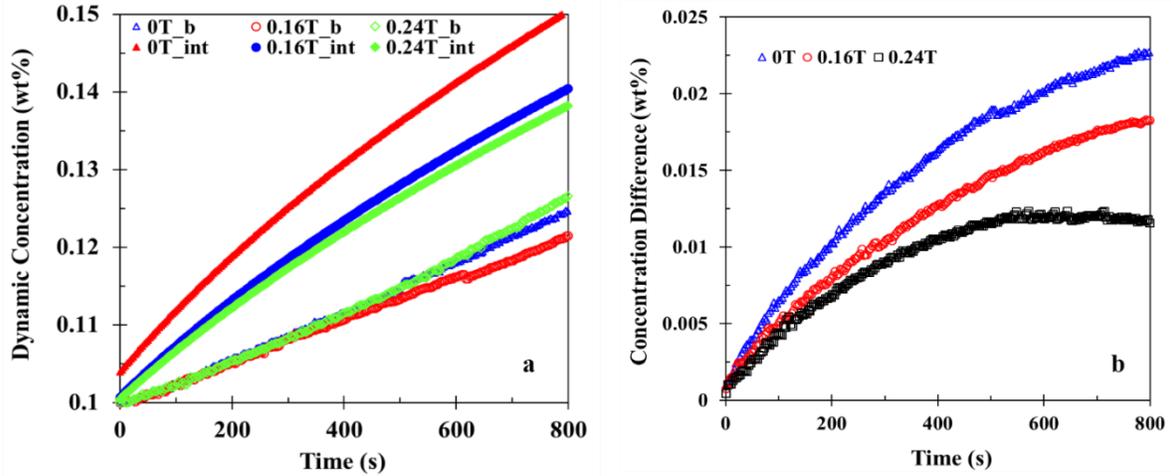

**Figure 9:** (a) The dynamic bulk and interfacial concentration nanoparticles in 0.1 wt. % $Fe_2O_3$ ferrofluid drop during evaporation under different field strengths, (b) the difference between the two concentrations corresponding to the cases in (a).

A scaling analysison the species transport within the drop is presented (refer supplementary for detailed discussion). The species transport balance is expressed as

$$\dot{m} = DA\frac{\Delta C_s}{R} + u_{c,s}A\Delta C_s + u_{f,s}A\Delta C_s \qquad (12)$$

where $\dot{m}$, $D$, and $\Delta C_s$ represent the mass loss rate due to evaporation, coefficient of diffusion of the nanoparticles in water, and the difference between the instantaneous bulk and interfacial concentrations within the droplet. $u_{c,s}$ is the circulation velocity due to solutal gradient only at zero field and $u_{f,s}$ is the ferro-advection velocity induced by magneto-solutal effect.

The expression for circulation velocity due to the solutal gradient is expressed as [31]

$$u_{c,s} = \frac{\sigma_c \Delta C_s}{\mu} \tag{13}$$

where $\sigma_c$ represents the rate of change of surface tension with respect to the concentration of nanoparticles. Substituting the expression of field induced circulation velocity $u_{f,s}$ (eqn. 4) and $u_{c,s}$ (eqn. 13), the eqn. 12 becomes

$$\rho R \dot{R} = D \Delta C_s + \frac{\sigma_c}{\mu} R (\Delta C_s)^2 + Ha \Delta C_s R \left( \frac{\rho u_{c,s}}{R \mu} \right)^{1/2} \tag{14}$$

For stable advection $1 + Ma_s \sim Ma_s$, and the expression becomes

$$\frac{\rho R \dot{R}}{D \Delta C_s} = Ma_s + \sqrt{HaMa_s} \sqrt{HaSc} = Ma_s + \sqrt{Ma_{s,m} Sc_m} \tag{15}$$

where $Ma_s = \frac{R}{\mu} \left( \frac{\sigma_c \Delta C_s}{D} \right)$ is the solutal Marangoni number and $Sc$ is the Schmidt number. In eqn. 15, $Ma_{s,m}$ and $Sc_m$ represent the magneto-solutal Marangoni number and the magnetic Schmidt number.

The percentage contribution of each component of the species balance (eqn. 12) in case of 0.1 wt% $Fe_2O_3$ based ferrofluid is illustrated in fig. 7 (b). It is noted that the magneto-solutal advection or soluto-ferro-hydrodynamic component is much pronounced compared to the thermal counterpart (except at high field strength of 0.24 T). The scaling thereby shows that the solutal-ferro-hydrodynamics component is the major contributor towards the internal advection dynamics observed in field environment. However, since the thermo-ferro-hydrodynamic component is also present, additional analysis is essential to infer the dominant mode.

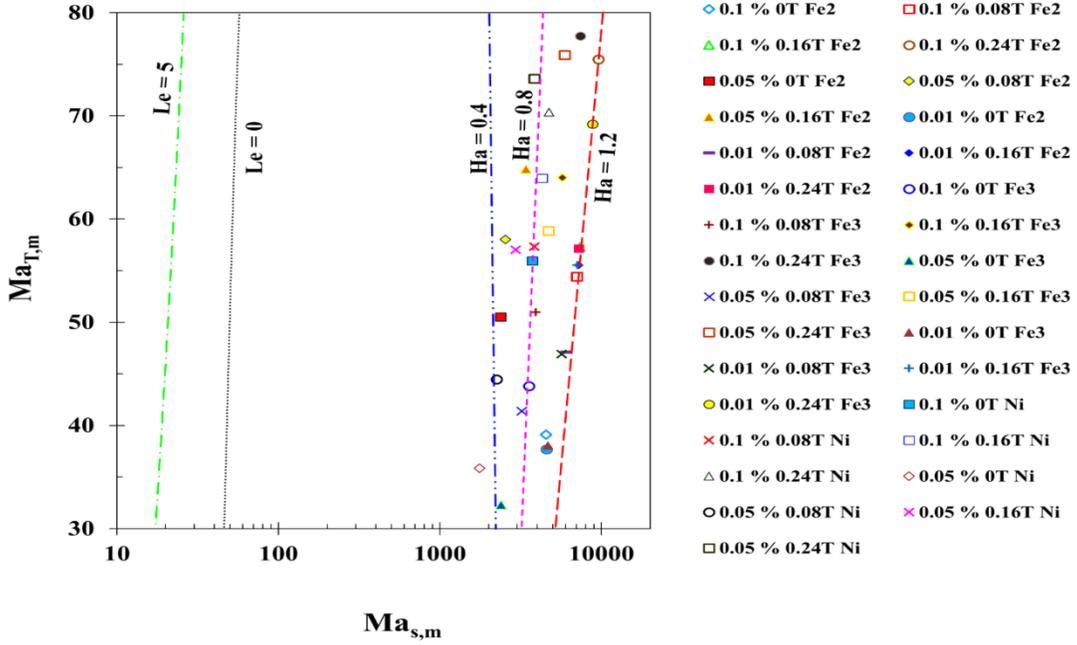

**Figure 10:** Phase map of $Ma_{T,m}$ and $Ma_{s,m}$. The lines indicate different iso-Le stability conditions (Joo[41]) and iso-Ha conditions.

The $Ma_{T,m}$ has been compared against the magneto-solutal Marangoni numbers ($Ma_{s,m}$) in fig. 10, for different magnetic field cases, in a phase map. The stability of advection criterion (Joo [41]) is indicated iso-Lewis numbers ($Le$) lines. The criterion states that any point residing to the right of the $Le=0$ line exhibits stable internal advection. This is noted to be true for all the points, with the points shifting further away from the $Le=0$ line with increase in the $Ha$. The thermal stability map however illustrated that the thermo-ferro-advection is at best partially stable for all cases. This signifies that the soluto-ferro-advection is the stable advection component and is the genesis of the observed internal advection behavior. The iso-Ha lines reveal that with increasing $Ha$, the points shift towards higher $Ma_{s,m}$ with reduction in the $Ma_{T,m}$.

## 3.F. Predicting the ferro-hydrodynamic advection velocity and evaporation rates for the ferrofluid droplets

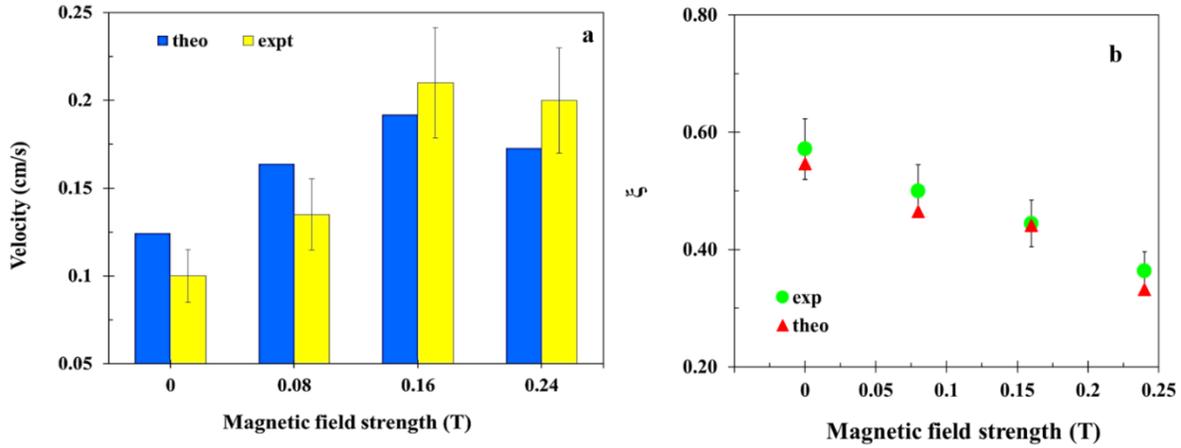

**Figure 11:** (a) Comparison of the predicted internal circulation velocity against experimental observations for 0.01 wt. % $Fe_2O_3$ ferrofluids, (b) comparison between theoretical ratio of evaporation rates of water and ferrofluid droplet with the experimental observations.

Fig. 11 (a) illustrates the comparison of theoretically predicted and the experimentally observed internal circulation velocities. Since the analysis shows the magneto-solutal effect to be the dominant mechanism behind the internal ferro-advection, the effective magneto-solutal Marangoni number has been used to determine the theoretical velocities of advection. The experimental velocities are the spatio-temporally averaged velocities obtained from velocimetry. The predictability of the scaled velocity is good for all magnetic fields. However, at 0.24 T, reduction in the predicted velocity value is noted. At 0.24 T the magneto-thermal component is also fairly strong (fig. 7 a) and it is plausible that only considering the magneto-solutal advection velocity leads to certain errors in the prediction at high field strengths. The overall predictability of the model shows that the proposed scaling analysis is valid and may be used to predict droplet evaporation kinetics.

To predict the evaporation rates of the ferrofluid droplets, a scaling model considering the Stefan flow [42] between the vapor diffusion layer shrouding the droplet and the ambient has been proposed (refer supplementary information for derivation). The proposed scaling model relates the evaporation rates of the water and ferrofluid (indicated by subscript '*p*') droplets as

$$\xi = \frac{\mu_l u_c t_p}{\mu_{l,p} u_{c,p} t} \tag{16}$$

where $\xi$ is the ratio of the evaporation rates ($k/k_p$). The $\mu$, $u_c$ and $t$ represent the viscosity, circulation velocity and droplet life times of the respective fluids. Fig. 11 (b) illustrates the comparison of the model predictions against the experimental observations. Agreement between the experiment and scaling model predictions further illustrates that the role of internal ferro-hydrodynamics and its influence on the external vapor layer is the underlying mechanism behind the observations.

Before probing into the improved evaporation rate due to the magnetic field, insight on the improved evaporation of the ferrofluids (at zero field) compared to water is essential. Evaporation kinetics of the colloidal ferrofluid is governed by many factors apart from the typical diffusion driven evaporation [23] in case of pure fluid droplets. Recent studies have reported the presence of soluto-thermal Marangoni advection cells inside evaporating pendant droplets of colloids [23]. The evaporating colloid droplets exhibit internal advection currents, which may also possess certain temporal periodicity. The ferrofluid droplets are not an exception to this, as evident from the flow visualization study, which shall be discussed subsequently. Some recent studies have explored the physics behind such advection in saline fluids [31, 35, 37] and the present observations can also be explained on similar grounds. The internal hydrodynamics shears the droplet-vapor interface, leading to entrainment of the ambient air, which replenishes the species boundary layer at the droplet-vapor interface. This replenishment leads to enhanced evaporation following the diffusion kinetics model.

However, the strength of advection in colloid droplets is not significantly high (at zero field) in order to improve the evaporation as noted in fig. 2. The role of the particle transport within the droplet also requires analysis. In a colloid droplet, the population of the particles at the air-water interface and that at the bulk is governed by the adsorption-desorption kinetics of the particles with respect to the base fluid [10, 19]. The Peclet number (*Pe*) is a measure of relative importance of convective transport to diffusive transport. Therefore, at low *Pe*, diffusion driven transport dominates and particle motion within the droplet is strongly controlled by the concentration gradient of the particulate phase. The particle diffusivity *D* can be correlated to the particle concentration Φ from the generalized Stokes-Einstein equation as (Yiantsios and Higgins [43])

$$D = D_0 \left(1 - \Phi\right)^{6.55} \frac{d}{d\Phi}\left(\frac{1.85\Phi}{0.64 - \Phi}\right) \tag{17}$$

The $D_0$ is the Einstein-Brownian diffusivity, expressed as $D_0 = \frac{k_B T}{6\pi\mu_0 a}$, where $k_B$ is the Boltzmann constant, $T^*$ is a characteristic temperature (taken to be the saturation temperature $T_{sat}$), and a represents the particle radius. Initially, as the particle concentration is increased, the diffusivity decreases due to interparticle hindrance effects (such as steric, electrostatic, magnetic (in the present case), etc.). However, the diffusivity increases with further increase of concentration, and ultimately diverges at a maximum concentration, $\Phi_m$. The diffusive particle transport, governed by the *Pe*, plays important role in determining the particle dynamics and migration to and from the droplet-air interface during evaporation [44]. The particles residing predominantly at the air-liquid interface tend to form a thin film or skin as the liquid molecules leave the interface due to evaporation. With time, as the concentration of the particles increase within the droplet, the propensity of formation of the skin enhances. This behavior is typically notable in case of sessile droplets on superhydrophobic surfaces, where buckling instability of the film confirms the formation and existence of such a film [45, 46]. Additionally, the hydrophilicity of the particles, the nature of interaction between particle and fluid, presence of surface active agents also influence the interfacial adsorption-desorption mechanism and can affect the skin formation [19].

The particle film-front at the interface is theorized to lead to the colloidal evaporation phenomena. The particle skin layer developed at the interface enhances evaporation by localized transpiration at the liquid-air interface of the droplet, by the action of the porous capillary wicking through the particle layers. The numerous available sites between the particle and the fluid medium draws more liquid front to the interface through nanoscale wicking or capillarity [47, 48] and this transpiration enhance the overall rate of evaporation [49]. In addition, the ferrofluid droplet shows mild internal advection (discussed subsequently), and thereby the diffusivity of the particles is spatio-temporally variant. In case of non-constant diffusivity, the particle layering at the interface is a direct function of the particle concentration only [44], which explains the improved evaporation rate of the ferrofluid (zero-field) with increasing particle concentration.

The interfacial layering and particle skin formation at the droplet-air interface in the presence of internal advection currents may augment the transport phenomena. The advection currents reduce the thickness of the interfacial layering, and the mixing effect within the droplet aids the localized nanoscale capillary wicking behavior. This can further enhance the evaporation rate in conjunction to the interfacial shear induced replenishment of the diffusion layer (as discussed earlier). Also, the inclusion of nanoparticles reduces the effective heat of vaporization of a colloid as [50]

$$\left(\rho h_{fg}\right)_{nf} = (1-\Phi)\rho h_{fg} + \left(\frac{T_{bf}}{T_{b,s}}\right)\Phi \rho_s h_{fg,s} \tag{18}$$

In eqn. 18, 'nf' represents the ferrofluid, 's' for the nanoparticles and no subscripts represents the basefluid. $T_{bf}$ is the boiling point of the basefluid and $\Phi$ is the concentration of the ferrofluid. Based on literature report [51], the enthalpy of vaporization can be expressed as

$$h_{fg} = \frac{TM\sigma^{1/4}}{[P]\rho_f \rho_g}\left(\frac{dP}{dT}\right)_{sat} \tag{19}$$

The parameter ($[P] = \frac{\sigma^{1/4}M}{\rho_f}$) is an empirical constant depending on the surface tension, density, and molecular weight of the fluid. The studies show that with ~ 2% volume fraction of the particles, the heat of vaporization of the colloids can be changed by ~ 30%.

The vapor mass fraction (*Y*) shrouding the droplet is directly proportional to the mole fraction (*x*), which is directly related to the vapor pressure at the droplet-air interface. This is correlated from the Clausius–Clapeyron equation as

$$\frac{dP}{P} = \frac{h_{fg}}{R/MW} \frac{dT}{T^2} \tag{20}$$

Hence, reduction in the enthalpy of vaporization enhances the mole fraction and the vapor mass fraction around the droplet. This directly affects the Spalding mass transfer number ($B_M$) as

$$B_M = \frac{Y_s - Y_\infty}{1 - Y_s} \tag{21}$$

where $Y_s$ is the mass fraction of the vapor at the droplet surface and $Y_\infty$ is the far field mass fraction of the vapor. The increase in $B_M$ results in enhancement of the evaporation constant ($k$) as.

$$\frac{dD^2}{dt} = \frac{-8\rho_g D_v}{\rho_l} \ln(1 + B_M) = -k \tag{22}$$

The improved evaporation of the ferrofluids is thus due to conglomeration of the particle film formation and change in the heat of vaporization due to presence of the particles. The evaporation rate increases with increasing concentration of particles and saturates at higher concentrations (fig. 2(b) and (d), 3(b) and 4 (b)). At higher concentrations, the viscosity of the ferrofluid enhances and it opposes the evaporation improving factors, such as particle skin formation, and internal advection.

## 4. Conclusions

The present article discusses the physics and mechanics of evaporation of pendent, aqueous ferrofluid droplets and modulation of the same by external magnetic field. The study uses detailed experiments to investigate the evaporation dynamics of pendent droplets of stable aqueous ferrofluids, and the modulation in dynamics due to the presence of external magnetic field. Optical imaging, infrared imaging, particle image velocimetry and interfacial property measurements are done in details to shed insight on the physics of the problem. The major conclusions from the study are as follows:

- Aqueous ferrofluid droplets evaporate at faster rates than water droplets. The improved evaporation behavior of such colloidal droplets has been explained based on the modulated

- heat of vaporization due to the nanoparticles, and the formation of thin particle shell structure which promotes faster evaporation by capillary action.
- In the presence of magnetic field, the evaporation rate of the ferrofluid droplets enhance further and this depends on the magnetic field strength, concentration of the magnetic nanoparticles, magnetic moment of the nanoparticles, etc.
- Changes in the surface tension, and classical diffusion driven evaporation models are insufficient towards explaining the improved evaporation. Internal probing using particle image velocimetry reveals changed internal advection dynamics within the droplet due to the field. Infrared imaging shows changed thermal states within the droplet due to evaporation in presence of field. Mapping the dynamic surface tension reveals that solutal advection is also present within the ferrofluid droplet.
- Two major advection modes, viz. the magneto-thermal and the magneto-solutal advection kinetics are deemed responsible towards the internal hydrodynamics. A scaling analysis is proposed and the magneto-solutal advection is noted to be the more dominant mechanism. The internal advection velocities are also predicted accurately from the analysis.
- It is shown that the modulation of the associated Stefan flow by the interfacial shear (caused by the internal advection) is responsible for the enhanced evaporation. A scaling model proposed along the same lines is able to predict the enhanced evaporation rates accurately.

The present findings may be of importance and have strong implications towards understanding transport phenomena in microhydrodynamics systems employing magnetic fluids.

## Acknowledgements

PD thanks the Department of Mechanical Engineering of IIT Ropar and IIT Kharagpur for partial financial support towards the present work. Also partial funding through the ISIRD project (IITRPR/Research/193) by IIT Ropar is acknowledged.

## Conflict of interest

The authors declare having no conflicts of interest with any individual or agency with respect to this work.

# References

1. G. A McConaghy, and B. A. Finlayson. "Surface tension driven oscillatory instability in a rotating fluid layer." *Journal of Fluid Mechanics* 39, no. 1 (1969): 49-55.
2. C. Golia, and A.Viviani. "Marangoni-buoyant boundary layers." (1985): 29-35.
3. A. M. Cazabat, F. Heslot, S. M. Troian, and P. Carles. "Fingering instability of thin spreading films driven by temperature gradients." *Nature* 346, no. 6287 (1990): 824.A
4. G. M. Faeth, "Evaporation and combustion of sprays." *Progress in Energy and Combustion Science* 9, no. 1-2 (1983): 1-76.
5. C. H. Chiang, M. S. Raju, and W. A. Sirignano. "Numerical analysis of convecting, vaporizing fuel droplet with variable properties." *International journal of heat and mass transfer* 35, no. 5 (1992): 1307-1324.
6. William A. Sirignano, "Fuel droplet vaporization and spray combustion theory." *Progress in Energy and Combustion Science* 9, no. 4 (1983): 291-322.
7. T. Kawase, H.Sirringhaus, R. H. Friend, and T.Shimoda. "Inkjet printed via-hole interconnections and resistors for all-polymer transistor circuits." *Advanced Materials* 13, no. 21 (2001): 1601-1605.
8. H. Hu, and R. G. Larson. "Analysis of the effects of Marangoni stresses on the microflow in an evaporating sessile droplet." *Langmuir* 21, no. 9 (2005): 3972-3980.
9. A. R. Harikrishnan, S. K. Das, P. K. Agnihotri, and P.Dhar. "Particle and surfactant interactions effected polar and dispersive components of interfacial energy in nanocolloids." *Journal of Applied Physics* 122, no. 5 (2017): 054301.
10. A. R. Harikrishnan, P.Dhar, P. K. Agnihotri, S.Gedupudi, and S. K. Das. "Wettability of complex fluids and surfactant capped nanoparticle-induced quasi-universal wetting behavior." *The Journal of Physical Chemistry B* 121, no. 24 (2017): 6081-6095.
11. K. Du, E.Glogowski, T.Emrick, T. P. Russell, and A. D. Dinsmore. "Adsorption energy of nano-and microparticles at liquid− liquid interfaces." *Langmuir* 26, no. 15 (2010): 12518-12522.
12. B. He, and F.Duan. "Evaporation and convective flow pattern of a heated pendant silicone oil droplet." *International Journal of Heat and Mass Transfer* 85 (2015): 910-915.

13. R. Savino, and S. Fico. "Transient Marangoni convection in hanging evaporating drops." *Physics of Fluids* 16, no. 10 (2004): 3738-3754.
14. D. K. Mandal, and S.Bakshi. "Internal circulation in a single droplet evaporating in a closed chamber." *International Journal of Multiphase Flow* 42 (2012): 42-51.
15. G. A. E. Godsave, "Studies of the combustion of drops in a fuel spray—the burning of single drops of fuel." In *Symposium (international) on combustion*, vol. 4, no. 1, pp. 818-830. Elsevier, 1953.
16. V. A. Kuz, "Fluid dynamic analysis of droplet evaporation." *Langmuir* 8, no. 11 (1992): 2829-2831.
17. C. H. Chon, S. Paik, J. B. Tipton, and K. D. Kihm. "Effect of nanoparticle sizes and number densities on the evaporation and dryout characteristics for strongly pinned nanofluid droplets." *Langmuir* 23, no. 6 (2007): 2953-2960.
18. T. AH Nguyen, and Anh V. Nguyen. "Increased evaporation kinetics of sessile droplets by using nanoparticles." *Langmuir* 28, no. 49 (2012): 16725-16728.
19. A. R. Harikrishnan, P.Dhar, P. K. Agnihotri, S.Gedupudi, and S. K. Das. "Effects of interplay of nanoparticles, surfactants and base fluid on the surface tension of nanocolloids." *The European Physical Journal E* 40, no. 5 (2017): 53.
20. H. S. Wi, S. Cingarapu, Ken J. Klabunde, and B. M. Law. "Nanoparticle adsorption at liquid–vapor surfaces: influence of nanoparticle thermodynamics, wettability, and line tension." *Langmuir* 27, no. 16 (2011): 9979-9984.
21. W. J.Gerken, A. V. Thomas, N.Koratkar, and Matthew A. Oehlschlaeger. "Nanofluid pendant droplet evaporation: experiments and modeling." *International Journal of Heat and Mass Transfer* 74 (2014): 263-268.
22. R-H Chen, T. X. Phuoc, and D. Martello. "Effects of nanoparticles on nanofluid droplet evaporation." *International Journal of Heat and Mass Transfer* 53, no. 19-20 (2010): 3677-3682.
23. A. R. Harikrishnan, P.Dhar, S.Gedupudi, and Sarit K. Das. "Oscillatory solutothermal convection-driven evaporation kinetics in colloidal nanoparticle-surfactant complex fluid pendant droplets." *Physical Review Fluids* 3, no. 7 (2018): 073604.
24. A. R. A. Fattah, S. Ghosh, and I. K. Puri. "Printing microstructures in a polymer matrix using a ferrofluid droplet." *Journal of Magnetism and Magnetic Materials* 401 (2016): 1054-1059.

25. A. Ray, V. B. Varma, Z. Wang, Z. Wang, P. J. Jayaneel, N. M. Sudharsan, and Raju V. Ramanujan. "Magnetic droplet merging by hybrid magnetic fields." *IEEE Magnetics Letters* 7 (2016): 1-5.
26. Z-G Guo, F. Zhou, J-C Hao, Y-M Liang, W-M Liu, and Wilhelm TS Huck. ""Stick and slide" ferrofluidic droplets on superhydrophobic surfaces." *Applied Physics Letters* 89, no. 8 (2006): 081911.
27. Fred K. Wohlhuter, and O. A. Basaran. "Shapes and stability of pendant and sessile dielectric drops in an electric field." *Journal of Fluid Mechanics* 235 (1992): 481-510.
28. O. A. Basaran, and Fred K. Wohlhuter. "Effect of nonlinear polarization on shapes and stability of pendant and sessile drops in an electric (magnetic) field." *Journal of Fluid Mechanics* 244 (1992): 1-16.
29. S. N. Reznik, A. L. Yarin, A. Theron, and E. Zussman. "Transient and steady shapes of droplets attached to a surface in a strong electric field." *Journal of Fluid Mechanics* 516 (2004): 349-377.
30. N. Havard, F.Risso, and PhTordjeman. "Breakup of a pendant magnetic drop." *Physical Review E* 88, no. 1 (2013): 013014.
31. V. Jaiswal, R. K.Dwivedi, A. R. Harikrishnan, and P.Dhar. "Magnetohydrodynamics-and magnetosolutal-transport-mediated evaporation dynamics in paramagnetic pendant droplets under field stimulus." *Physical Review E* 98, no. 1 (2018): 013109.
32. C. A. Schneider, W. S. Rasband, and K. W. Eliceiri. "NIH Image to ImageJ: 25 years of image analysis." *Nature methods* 9, no. 7 (2012): 671.
33. W. Thielicke, and E.Stamhuis. "PIVlab–towards user-friendly, affordable and accurate digital particle image velocimetry in MATLAB." *Journal of Open Research Software* 2, no. 1 (2014).
34. B. Abramzon, and W. A. Sirignano. "Droplet vaporization model for spray combustion calculations." *International journal of heat and mass transfer* 32, no. 9 (1989): 1605-1618.
35. P. Dhar, V. Jaiswal, and A. R. Harikrishnan. "Electromagnetic field orientation and characteristics governed hydrodynamics within pendent droplets." Physical Review E 98, no. 6 (2018): 063103.

36. V. Jaiswal, and P.Dhar. "Interplay of electro-thermo-solutal advection and internal electrohydrodynamics governed enhanced evaporation of droplets." *Proceedings of the Royal Society A* 475, no. 2225 (2019): 20190046.

37. V. Jaiswal, A. R. Harikrishnan, G.Khurana, and P.Dhar. "Ionic solubility and solutal advection governed augmented evaporation kinetics of salt solution pendant droplets." *Physics of Fluids* 30, no. 1 (2018): 012113.

38. D. A. Nield, "Surface tension and buoyancy effects in cellular convection." *Journal of Fluid Mechanics* 19, no. 3 (1964): 341-352.

39. S. H. Davis, "Buoyancy-surface tension instability by the method of energy." *Journal of Fluid Mechanics* 39, no. 2 (1969): 347-359.

40. N. Rudraiah, R. M. Barron, M. Venkatachalappa, and C. K. Subbaraya. "Effect of a magnetic field on free convection in a rectangular enclosure." *International Journal of Engineering Science* 33, no. 8 (1995): 1075-1084.

41. S. W. Joo, "Marangoni instabilities in liquid mixtures with Soret effects." *Journal of Fluid Mechanics* 293 (1995): 127-145.

42. S. R. Turns, *An introduction to combustion*. Vol. 499. New York: McGraw-hill, 1996.

43. S. G. Yiantsios, and B. G. Higgins. "Marangoni flows during drying of colloidal films." *Physics of Fluids* 18, no. 8 (2006): 082103.

44. K. L. Maki, and S. Kumar. "Fast evaporation of spreading droplets of colloidal suspensions." *Langmuir* 27, no. 18 (2011): 11347-11363.

45. H. Xu, S.Melle, K.Golemanov, and Gerald Fuller. "Shape and buckling transitions in solid-stabilized drops." *Langmuir* 21, no. 22 (2005): 10016-10020.

46. P. J. Yunker, M.Gratale, M.A. Lohr, T. Still, T. C. Lubensky, and A. G. Yodh. "Influence of particle shape on bending rigidity of colloidal monolayer membranes and particle deposition during droplet evaporation in confined geometries." *Physical review letters* 108, no. 22 (2012): 228303.

47. X. Chen, J. Chen, X. Ouyang, Y. Song, R.Xu, and P. Jiang. "Water droplet spreading and wicking on nanostructured surfaces." *Langmuir* 33, no. 27 (2017): 6701-6707.

48. R. Dufour, P. Brunet, M.Harnois, R.Boukherroub, V.Thomy, and V.Senez. "Zipping effect on omniphobic surfaces for controlled deposition of minute amounts of fluid or colloids." *Small* 8, no. 8 (2012): 1229-1236.

49. N. Fries, K. Odic, M. Conrath, and M. Dreyer. "The effect of evaporation on the wicking of liquids into a metallic weave." *Journal of colloid and interface science* 321, no. 1 (2008): 118-129.
50. M. Mehregan, and M.Moghiman. "Propose a correlation to approximate nanofluids' enthalpy of vaporization—a numerical study." *International Journal of Materials, Mechanics and Manufacturing* 2, no. 1 (2014): 73-6.
51. S. Krishnamurthy, P. Bhattacharya, P. E. Phelan, and R. S. Prasher. "Enhanced mass transport in nanofluids." *Nano letters* 6, no. 3 (2006): 419-423.